\documentclass{epl}
\usepackage{epsfig}		

\def\0{\mbox{\tiny $0$}}
\def\1{\mbox{\tiny $1$}}
\def\2{\mbox{\tiny $2$}}
\def\3{\mbox{\tiny $3$}}
\def\4{\mbox{\tiny $4$}}
\def\5{\mbox{\tiny $5$}}
\def\6{\mbox{\tiny $6$}}
\def\7{\mbox{\tiny $7$}}
\def\8{\mbox{\tiny $8$}}
\def\9{\mbox{\tiny $9$}}

\def\f14{\mbox{\tiny $\frac{1}{4}$}}
\def\infm{\mbox{\tiny $-\infty$}}
\def\infp{\mbox{\tiny $+\infty$}}

\def\ii{\mbox{\tiny $i$}}
\def\z{\mbox{\tiny $z$}}

\def\mi{\mbox{\tiny $-$}}

\title{Additional time-dependent phase in the flavor-conversion formulas}
\shorttitle{Time-dependent phase in the flavor-conversion formulas}

\author{A. E. Bernardini\inst{1}}
\institute{
  \inst{1} Instituto de F\'{\i}sica Gleb Wataghin, UNICAMP,\\
  PO Box 6165, 13083-970, Campinas, SP, Brasil.}

\pacs{02.30.Mv}{First pacs description}
\pacs{03.65.-w}{Second pacs description}
\pacs{11.30.Hv}{Third pacs description}

\begin{document}

\maketitle

\begin{abstract}
In the framework of intermediate wave-packets for treating flavor oscillations, we quantify the modifications which appear when we assume a strictly peaked momentum distribution  and consider the second-order corrections in a power series expansion of the energy.
By following a sequence of analytic approximations, we point out that an extra time-dependent phase is merely the residue of second-order corrections.
Such phase effects are usually ignored in the relativistic wave-packet treatment, but they do not vanish non-relativistically and can introduce some small modifications to the oscillation pattern even in the ultra-relativistic limit.
\end{abstract}

Over recent years, the quantum mechanics of oscillations \cite{Zub98,Alb03,Vog04} has experienced
much progress on the theoretical front \cite{Beu03}, in particular, not only in phenomenological pursuit of a more refined flavor
conversion formula \cite{Giu98,Zra98,Ber05} which, sometimes, deserves a special attention,
but also in efforts to give the theory a formal structure within quantum field formalism \cite{Bla95,Giu02B,Bla03}.
Under the point of view of a first quantized theory,
the flavor oscillation phenomena discussed in terms of
the {\em intermediate} wave-packet approach \cite{Kay81} eliminates the most controversial
points rising up with the {\em standard} plane-wave formalism \cite{Kay89,Kay04}.
In fact, wave-packets describing propagating mass-eigenstates
guarantees the existence of a coherence length \cite{Kay81}, avoids the ambiguous approximations in the plane
wave derivation of the phase difference \cite{DeL04} and, under particular conditions of
minimal {\em slippage}, recovers the oscillation probability given by the {\em standard}
plane wave treatment.
Otherwise, strictly speaking, the {\em intermediate} wave-packet formalism can also be refuted, for example, 
in the context of neutrino oscillation since such oscillating particles are neither prepared nor observed \cite{Beu03} in this case.
Some authors suggest the calculation of a transition probability between
the observable particles involved in the production and detection
process in the so-called {\em external}
wave-packet approach \cite{Giu02B,Beu03,Ric93}: the oscillating
particle, described as an internal line of a Feynman diagram by a
relativistic mixed scalar propagator, propagates between the
source and target (external) particles represented by wave
packets.
It can be demonstrated \cite{Beu03}, however, that the overlap function of
the incoming and outgoing wave-packets
in the {\em external} wave-packet model is mathematically equivalent
to the wave function of the propagating mass-eigenstate in the
{\em intermediate} wave-packet formalism.
Thus, as a preliminary investigation concerning with the existence of
an extra time-dependent phase added to the {\em standard} oscillation term $\frac{\Delta m^{\2} \,t}{2\, p_{\0}}$ \cite{Kay04}, 
we avoid the field theoretical methods in detriment to a clearer 
treatment with {\em intermediate} wave-packets
which commonly simplifies the
understanding of physical aspects going with the oscillation
phenomena.

The main aspects of oscillation phenomena can be understood by
studying the two flavor problem. In addition, substantial
mathematical simplifications result from the assumption that the
space dependence of wave functions is one-dimensional ($z$-axis).
Therefore, we shall use these simplifications to calculate the
oscillation probabilities. In this context, the time evolution of
flavor wave-packets can be described by
\begin{eqnarray}
\Phi(z,t) &=& \phi_{\1}(z,t)\cos{\theta}\,\mbox{\boldmath$\nu_{\1}$} + \phi_{\2}(z,t)\sin{\theta}\,\mbox{\boldmath$\nu_{\2}$}\nonumber\\
          &=& \left[\phi_{\1}(z,t)\cos^{\2}{\theta} + \phi_{\2}(z,t)\sin^{\2}{\theta}\right]\,\mbox{\boldmath$\nu_\alpha$}+
		  \left[\phi_{\1}(z,t) - \phi_{\2}(z,t)\right]\cos{\theta}\sin{\theta}\,\mbox{\boldmath$\nu_\beta$}\nonumber\\
          &=& \phi_{\alpha}(z,t;\theta)\,\mbox{\boldmath$\nu_\alpha$} + \phi_{\beta}(z,t;\theta)\,\mbox{\boldmath$\nu_\beta$},
\label{0}
\end{eqnarray}
where {\boldmath$\nu_\alpha$} and {\boldmath$\nu_\beta$} are flavor-eigenstates
and {\boldmath$\nu_{\1}$} and {\boldmath$\nu_{\2}$} are mass-eigenstates.
The probability of finding a flavor state $\mbox{\boldmath$\nu_\beta$}$ at
the instant $t$ is equal to the integrated squared modulus of the
$\mbox{\boldmath$\nu_\beta$}$ coefficient
\begin{equation}
P(\mbox{\boldmath$\nu_\alpha$}\rightarrow\mbox{\boldmath$\nu_\beta$};t)=
\int_{_{-\infty}}^{^{+\infty}}\mbox{$dz$} \,\left|\phi_{\beta}\right|^{\2}
= 
\frac{\sin^{\2}{[2\theta]}}{2}
\left\{\, 1 - \mbox{\sc Fo}(t) \, \right\},
\label{1}
\end{equation}
where $\mbox{\sc Fo}(t)$ represents the
interference term given by
\begin{equation}
\mbox{\sc Fo}(t) = Re 
 \left[\, \int_{_{-\infty}}^{^{+\infty}}dz
\,\phi^{\dagger}_{\1}(z,t) \, \phi_{\2}(z,t) \, \right].
\label{2}
\end{equation}
Let us consider mass-eigenstate wave-packets given by
\begin{equation}
\phi_{\ii}(z,0) = \left(\frac{2}{\pi a^{\2}}\right)^{ \frac{1}{4}} \exp{\left[- \frac{z^{\2}}{a^{\2}}\right]} \exp{[i p_{\ii} \, z]},
\label{3}
\end{equation}
at time $t = 0$, where $i = 1,\, 2$.
The wave functions which describe their time evolution are
\begin{eqnarray}
\phi_{\ii}(z,t) =
\int_{_{-\infty}}^{^{+\infty}}\frac{dp_{\z}}{2 \pi} \,
\varphi(p_{\z} \mi p_{\ii}) \exp{\left[-i\,E^{(\ii)}_{p_{\z}}\,t +i \, p_{\z}
\,z\right]}
,
\label{4}
\end{eqnarray}
where
$E^{(\ii)}_{p_{\z}} = \left(p_{\z}^{\2} + m_{\ii}^{\2}\right)^{ \frac{1}{2}}$
and
$\varphi(p_{\z} \mi p_{\ii}) =  \left(2 \pi a^{\2} \right)^{ \frac{1}{4}} \exp{\left[- \frac{(p_{\z} \mi p_{\ii})^{\2}\,a^{\2}}{4}\right]}.$
In order to obtain the oscillation probability, we can calculate the interference term $\mbox{\sc Fo}(t)$
by solving the following integral
\begin{eqnarray}
\lefteqn{\int_{\infm}^{\infp}\mbox{$\frac{dp_z}{2 \pi} \,  \varphi(p_z \mi p_{ 1}) \varphi(p_z \mi p_{ 2})
\exp{[-i \, \Delta E_{(p_z)} \, t]}$} =}\nonumber\\ 
&& 
\exp{\left[ \frac{\mi(a \, \Delta{p})^{\2}}{8}\right]}
\int_{\infm}^{\infp}\mbox{$\frac{dp_z}{2 \pi}  \, \varphi^{\2}(p_z \mi p_{\0})\exp{[-i \, \Delta E_{(p_z)} \, t]}$},~~~ \label{6}
\end{eqnarray}
where we have changed the $z$-integration into a $p_{\z}$-integration
and introduced the quantities $\Delta p = p_{ \1} \mi p_{ \2}
,\,\, p_{\0} = \frac{1}{2}(p_{ \1} + p_{ \2})$ and $\Delta E_{p_{\z}} = E^{(\1)}_{p_{\z}} \mi E^{(\2)}_{p_{\z}}$
The oscillation term is
bounded by the exponential function of $a \, \Delta p$ at
any instant of time. Under this condition we could never observe a
{\em pure} flavor-eigenstate. Besides, oscillations are
considerably suppressed if $a \, \Delta p > 1$. A necessary
condition to observe oscillations is that $a \, \Delta p \ll 1$.
This constraint can also be expressed by $\delta p \gg \Delta p$
where $\delta p$ is the momentum uncertainty of the particle. The
overlap between the momentum distributions is indeed relevant only
for $\delta p \gg \Delta p$. Consequently, without loss of
generality, we can assume
\begin{equation}
\mbox{\sc Fo}(t) = Re 
\left\{\int_{_{-\infty}}^{^{+\infty}}\frac{dp_{\z}}{2
\pi}
 \, \varphi^{\2}(p_{\z} \mi p_{\0})\exp{[-i \, \Delta E_{p_{\z}} \, t]}\right\}
\label{9}.
\end{equation}

In the literature, this equation is often obtained by assuming two
mass eigenstate wave-packets described by the ``same'' momentum
distribution centered around the average momentum
$\bar{p} = p_{\0}$. This simplifying hypothesis  also guarantees
{\em instantaneous} creation of a {\em pure} flavor
eigenstate {\boldmath$\nu_\alpha$} at $t = 0$ \cite{DeL04}. In fact, for $
\phi_{\1}(z,0)=\phi_{\2}(z,0)$ we get from Eq.~(\ref{0})
\begin{equation}
\phi_{\alpha}(z,0,\theta) = \left(\frac{2}{\pi
a^{\2}}\right)^{\frac{1}{4}} \exp{\left[- \frac{z^{\2}}{a^{\2}}\right]}
\exp{[i  p_{\0} \,z]}
\nonumber\end{equation}
and
$\phi_{\beta}(z,0,\theta) =0$.
In order to obtain an expression for $\phi_{\ii}(z,t)$ by analytically solving the integral in Eq.~(\ref{4}) we firstly rewrite the energy $E^{(\ii)}_{p_{\z}}$ as
\begin{equation}
E^{(\ii)}_{p_{\z}} = E_{\ii} \left[1 + \frac{ p_{\z}^{\2} \mi p_{\0}^{\2}}{E_{\ii}^{\2}}\right]^{ \frac{1}{2}} = E_{\ii} \left[1 + \sigma_{\ii} \left(\sigma_{\ii} + 2 \mbox{v}_{\ii}\right)\right]^{ \frac{1}{2}},
\label{11}
\end{equation}
where $E_{\ii} = (m_{\ii}^{\2} + p_{\0}^{\2})^{ \frac{1}{2}}$, $\mbox{v}_{\ii} = \frac{ p_{\0}}{E_{\ii}}$ and 
$\sigma_{\ii} = \frac{ p_{\z} \mi  p_{\0}}{E_{\ii}}$.
The use of free {\em gaussian} wave
packets \cite{Coh77,Giu98,Giu02B,Beu03} is justified in non-relativistic quantum mechanics because, in most of the cases,
the calculations can be carried out
exactly for these particular functions.
The reason lies in the fact that the frequency components
of the mass-eigenstate wave-packets,
$E^{(\ii)}_{p_{\z}}= p_{\z}^{\2}/2 m_{\ii}$, modify the momentum
distribution into ``generalized'' {\em gaussian}, easily integrated by
well known methods of analysis. The term $ p_{\z}^{\2}$ in
$E^{(\ii)}_{p_{\z}}$ is then responsible for the variation in time
of the width of the mass-eigenstate wave-packets, the so-called
{\em spreading} phenomenon. In relativistic quantum mechanics the
frequency components of the mass-eigenstate wave-packets,
$E^{(\ii)}_{p_{\z}}=\sqrt{ p_{\z}^{\, \2} + m_{\ii}^{\2} }$, do not
permit  an immediate analytic integration. This difficulty,
however, may be remedied by assuming a sharply peaked
momentum distribution, i. e. $(a \, E_{\ii})^{-1}\sim\sigma_{\ii} \ll 1$.
Meanwhile, the integral in Eq.~(\ref{4}) can be {\em
analytically} solved only if we consider terms up to order
$\sigma_{\ii}^2$ in the series expansion.
In this case, we can conveniently
truncate the power series
\begin{eqnarray}
E^{(\ii)}_{p_{\z}} & = & E_{\ii} \left[1 + \sigma_{\ii} \mbox{v}_{\ii}  + \frac{\sigma_{\ii}^{\2}}{2}\left(1 - \mbox{v}_{\ii}^{\2} \right)\right] + \mathcal{O}(\sigma_{\ii}^{\3})
\nonumber\\  &
 \approx &
  E_{\ii} +  p_{\0} \sigma_{\ii} + \frac{m_{\ii}^{\2}}{2E_{\ii}} \sigma_{\ii}^{\2}.
\label{12}
\end{eqnarray}
and get an analytic expression for the
oscillation probability.
The zero-order term in the previous expansion, $E_{\ii}$, gives the
standard plane wave oscillation phase. The first-order term, $ p_{\0}
\sigma_{\ii}$, will be responsible for the {\em slippage}
 due to the different group velocities of the
mass-eigenstate wave-packets and represents a linear correction to
the standard oscillation phase \cite{DeL04}. Finally, the
second-order term, $\frac{m_{\ii}^2}{2E_{\ii}} \sigma_{\ii}^2$, which is a
(quadratic) secondary correction will give the well-known
{\em spreading} effects in the time propagation of the wave-packet and
will be also responsible for a {\em new} additional phase to be
computed in the final calculation. In the case of {\em gaussian}
momentum distributions for the mass-eigenstate wave-packets, these
terms can all be {\em analytically} quantified.
By substituting (\ref{12}) in Eq.~(\ref{4}) and changing the
$ p_{\z}$-integration into a $\sigma_{\ii}$-integration, we obtain the
explicit form of the mass-eigenstate wave-packet time evolution,
\begin{eqnarray}
\phi_{\ii}(z,t)
 &=& 
 \left[\frac{2}{\pi \,a^{\2}_{\ii}(t)}\right]^{ \frac{1}{4}}\exp{[-i\,(\theta_{\ii}(t, z) + E_{\ii} \, t -  p_{\0} \, z)]}
\exp{\left[-\frac{(z - \mbox{v}_{\ii} \,t)^{\2}}{a_{\ii}^{\2}(t)}\right]},
\label{13}
\end{eqnarray}
where
$\theta_{\ii}(t, z) = \left\{\frac{1}{2}\arctan{\left[\frac{2\,m_{\ii}^{\2}\, t}{a^{\2}\, E_{\ii}^{\3}}\right]} - \frac{2\, m_{\ii}^{\2}\, t} {a^{\2}\, E_{\ii}^{\3}}\,\frac{(z - \mbox{v}_{\ii} \,t)^{\2}}{a_{\ii}^{\2}(t)}\right\}$
and
$a_{\ii}(t) = a \left(1 + \frac{4\, m_{\ii}^{\4}}{a^{\4}\, E_{\ii}^{\6}}\,t^{\2}\right)^{ \frac{1}{2}}$.
The time-dependent quantities $a_{\ii}(t)$ and $\theta_{\ii}(t, z)$ contain all the physically significant information which arise from the second order term in the power series expansion (\ref{12}).
By solving the integral (\ref{9}) with the approximation (\ref{11})
and performing some mathematical manipulations, we obtain
\begin{equation}
\mbox{\sc Fo}(t) = \mbox{\sc Bnd}(t) \times \mbox{\sc Osc}(t),
\label{20}
\end{equation}
where we have factored the time-vanishing bound of the interference term given by
\begin{equation}
\mbox{\sc Bnd}(t) = \left[1 + \mbox{\sc Sp}^{\2}(t) \right]^{-\frac{1}{4}}
\exp{\left[-\frac{(\Delta \mbox{v} \, t)^{\2}}{2a^{\2}\left[1 + \mbox{\sc Sp}^{\2}(t)\right]}\right]}
\label{21}
\end{equation}
and the time-oscillating character of the flavor conversion formula given by
\begin{eqnarray}
\mbox{\sc Osc}(t) &=& Re \left\{\exp{\left[-i\Delta E \, t -i \Theta(t)\right]} \right\}\nonumber\\
&=& \cos{\left[\Delta E \, t + \Theta(t)\right]},
\label{22A}
\end{eqnarray}
where
\begin{equation}
\mbox{\sc Sp}(t) = \mbox{$\frac{t}{a^{\2}}\Delta\left(\frac{m^{\2}}{E^{\3}}\right) = \rho\, \frac{\Delta \mbox{v}\, t}{a^{\2} \,  p_{\0}}$}
\label{240}
\end{equation}
and
\begin{equation}
\Theta(t) = \mbox{$\left[\frac{1}{2}\arctan{\left[\mbox{\sc Sp}(t)\right]} - \frac{a^{\2} \,  p_{\0}^{\2}}{2 \rho^{\2}}\frac{\mbox{\sc Sp}^{\3}(t)}{\left[1 + \mbox{\sc Sp}^{\2}(t)\right]}\right]$}, 
\label{24A}
\end{equation}
with
$\rho =
 1 - \left[3 + \left(\frac{\Delta E}{\bar{E}}\right)^{\2}\right] \frac{ p_{\0}^{\2}}{\bar{E}^{\2}}$ and
$\bar{E} = \sqrt{E_{\1} \, E_{\2}}$.
The time-dependent quantities $\mbox{\sc Sp}(t)$ and $\Theta(t)$
carry the second order corrections and, consequently, the
{\em spreading} effect to the oscillation probability formula.
If $\Delta E \ll \bar{E}$, the parameter $\rho$ is limited by
the interval $[1,-2]$ and it assumes the zero value when $\frac{ p_{\0}^{\2}}{\bar{E}^{\2}} \approx \frac{1}{3}$.
Therefore, by considering increasing values of $ p_{\0}$,
from non-relativistic (NR) to ultra-relativistic (UR) propagation regimes,
and fixing $\frac{\Delta E}{a^{\2} \, \bar{E}^{\2}}$,
the time derivatives of $\mbox{\sc Sp}(t)$ and $\Theta(t)$ have their
signals inverted when $\frac{ p_{\0}^{\2}}{\bar{E}^{\2}}$ reaches the value $\frac{1}{3}$.
The {\em slippage} between the mass-eigenstate wave-packets is
quantified by the vanishing behavior of $\mbox{\sc Bnd}(t)$. 
In order to compare $\mbox{\sc Bnd}(t)$ with the correspondent
function without the second order corrections (without {\em spreading}),
\begin{equation}
\mbox{\sc Bnd}_{\mbox{\tiny $WS$}}(t) = \exp{\left[-\frac{(\Delta \mbox{v} \, t)^{\2}}{2a^{\2}}\right]},
\label{23A}
\end{equation}
we substitute ${\mbox{\sc Sp}(t)}$ given by the expression (\ref{22A})
in Eq.~(\ref{21}) and we obtain the ratio
\begin{equation}
\frac{\mbox{\sc Bnd}(t)}{\mbox{\sc Bnd}_{\mbox{\tiny $WS$}}(t)}
= 
\left[1 + \rho^{\2} \left(\frac{\Delta E \, t}{a^{\2} \, \bar{E}^{\2}}\right)^{\2} \right]^{-\frac{1}{4}}
\exp{\left[\frac{\rho^{\2} \,  p_{\0}^{\2} \,
\left(\Delta E \, t\right)^{\4}}
{2\,a^{\6} \, \bar{E}^{\8}\left[1 + \rho^{\2} \left(\frac{\Delta E \, t}{a^{\2} \, \bar{E}^{\2}}\right)^{\2}\right]}
\right]}.
\label{23AA}
\end{equation}
The NR limit is obtained by setting $\rho^{\2} = 1$ and $ p_{\0} = 0$ in Eq.~(\ref{23A}).
In the same way, the UR limit is obtained by setting $\rho^{\2} = 4$
and $ p_{\0} = \bar{E}$.
In fact, the minimal influence due to second order corrections occurs
when $\frac{ p_{\0}^{\2}}{\bar{E}^{\2}} \approx \frac{1}{3}$ ($\rho \approx 0$).
Returning to the exponential term of Eq.~(\ref{21}), we observe that the oscillation amplitude is more
relevant when $\Delta \mbox{v} \, t \ll a$.
It characterizes the {\em minimal slippage} between the mass-eigenstate
wave-packets which occur when the
complete spatial intersection between themselves starts to diminish
during the time evolution.
Anyway, under {\em minimal slippage} conditions, we always have
$\frac{\mbox{\sc Bnd}(t)}{ \mbox{\sc Bnd}_{\mbox{\tiny $WS$}}(t)} \approx 1$.

\begin{figure}
\hspace{-2.6cm}
\epsfig{file= 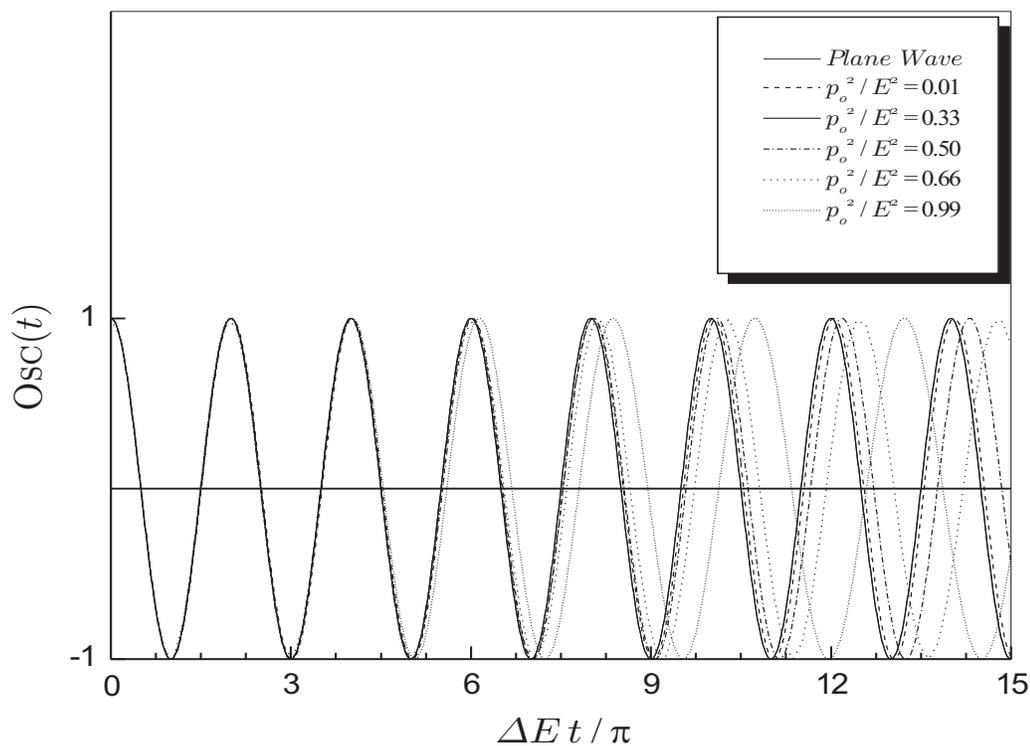, height= 13.0 cm, width= 17 cm}
\caption{The time-behavior of $\mbox{\sc Osc}(t)$ compared with the {\em standard} plane-wave oscillation given by $\cos{[\Delta E \, t]}$
for different propagation regimes.
The additional phase $\Theta(t)$ changes the oscillating character after some time of propagation.
The minimal deviation occurs for $\frac{p_{\0}^{\2}}{\bar{E}^{\2}} \approx \frac{1}{3}$ which is represented by a solid
line superposing the plane-wave case.}
\label{an4}
\end{figure}

\begin{figure}
\begin{center}
\epsfig{file= 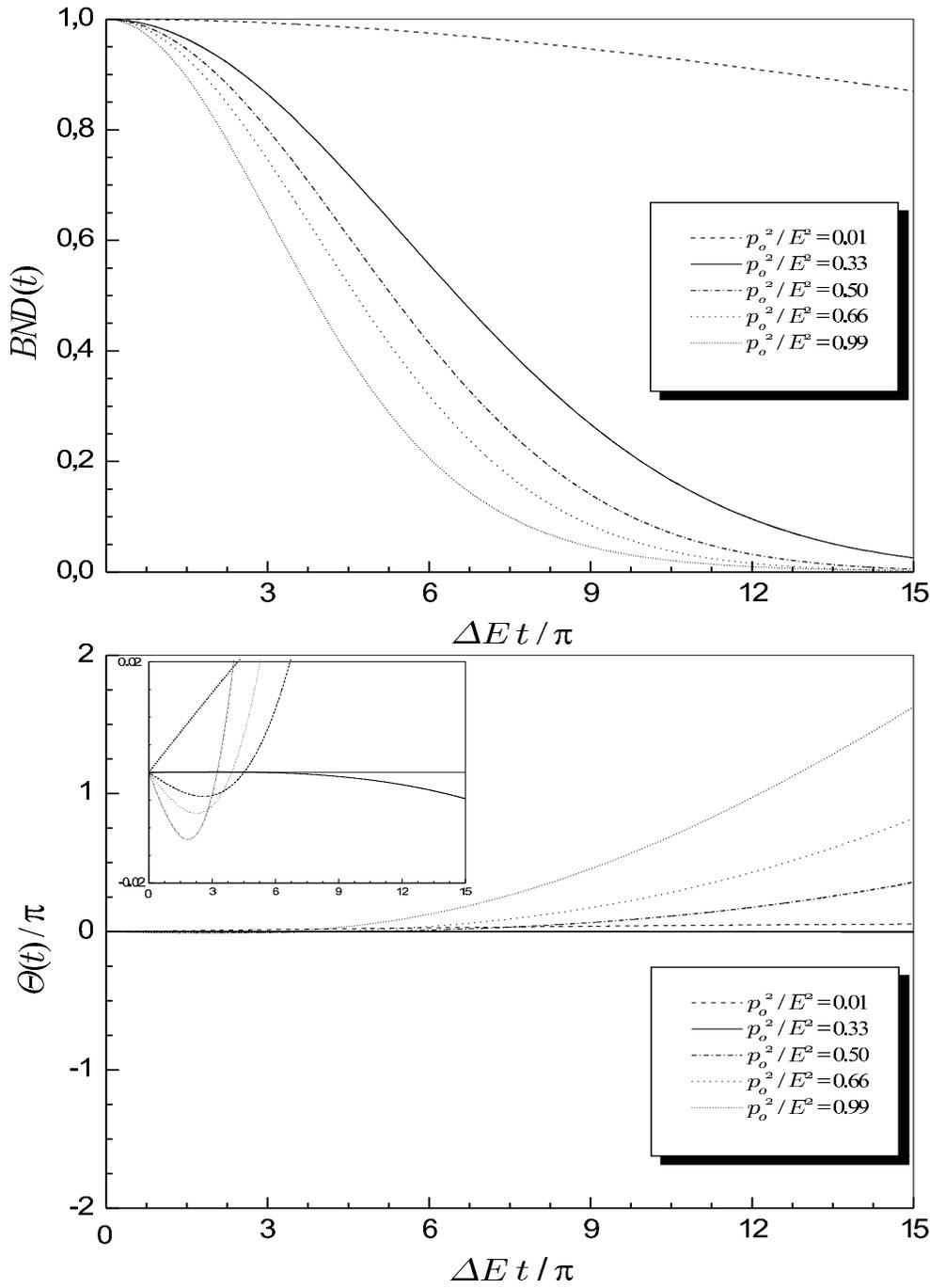, height= 18.0 cm, width= 13 cm}
\end{center}
\caption{The values assumed by $\Theta (t)$ are {\em  effective} while the interference term does not vanish.
In the upper box we can observe the behavior of $\mbox{\sc Bnd}(t)$ which determines the limit values effectively assumed by
$\Theta(t)$ for each propagation regime.
For relativistic regimes with $\frac{ p_{\0}^{\2}}{\bar{E}^{\2}} > \frac{1}{3}$, the function $\Theta(t)$ rapidly reaches its lower limit as we can observe in the small box above.
We have used $a \, \bar{E} = 10$.}
\label{an5}
\end{figure}

\begin{figure}
\hspace{-2.6cm}
\epsfig{file= 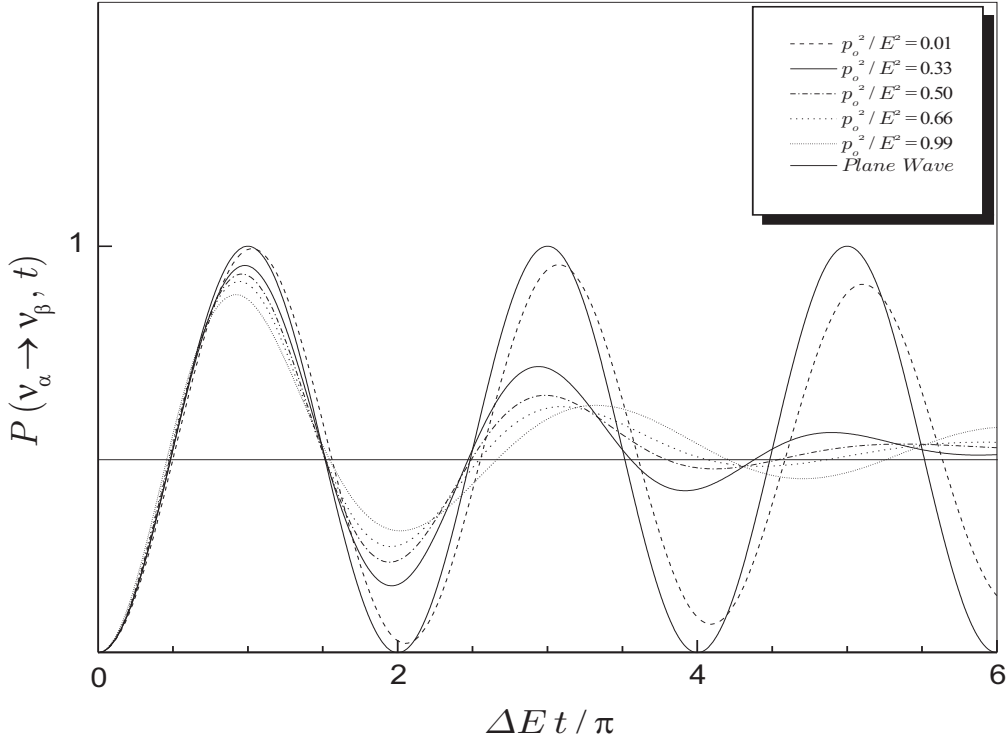, height= 13.0 cm, width= 17.0 cm}
\caption{Flavor conversion probability time dependence obtained with the introduction of
second-order corrections in the series expansion of the energy for a strictly peaked momentum distribution ($\mathcal{O}(\sigma_{\ii}^{\3})$).
By comparing with the PW predictions, depending on the propagation regime,
the additional time-dependent phase $\Delta \Phi(t) \equiv\, \Delta E \, t + \Theta(t)$ produces a delay/advance in the local maxima of flavor detection.
Phenomenologically, it can introduce small quantifiable deviations to the averaged detected values of neutrino oscillation parameters.
Essentially, it depends on the product of the wave-packet width $a$ by the averaged energy $\bar{E}$.}
\label{an8}
\end{figure}

The oscillating function $\mbox{\sc Osc}(t)$ of the interference
term $\mbox{\sc Fo}(t)$ differs from the {\em standard} oscillating
term, $ \cos{[\Delta E \, t]}$,
by the presence of the additional phase $\Theta(t)$
which is essentially a second order correction.
The modifications introduced by the additional phase $\Theta(t)$ are discussed in Fig.~\ref{an4}
where we have compared the time-behavior of $\mbox{\sc Osc}(t)$ to $\cos{[\Delta E \, t]}$ for different propagation regimes.
The bound {\em  effective} value assumed by $\Theta (t)$
is determined by the vanishing behavior of $\mbox{\sc Bnd}(t)$.
To illustrate this flavor oscillation behavior, we plot both the curves representing $\mbox{\sc Bnd}(t)$ and $\Theta(t)$ in Fig.~\ref{an5}.
We note the phase slowly changing in the NR regime.
The modulus of the phase $|\Theta(t)|$ rapidly reaches its upper limit when $\frac{ p_{\0}^{\2}}{\bar{E}^{\2}} > \frac{1}{3}$ and, after a certain time, it continues to evolve approximately linearly in time.
But, effectively, the oscillations rapidly vanishes.
By superposing the effects of $\mbox{\sc Bnd}(t)$ in Fig.~\ref{an5} and the oscillating character $\mbox{\sc Osc}(t)$ expressed in Fig.~\ref{an4}, we immediately obtain the flavor oscillation probability which is explicitly given by 
\begin{equation}
P(\mbox{\boldmath$\nu_\alpha$}\rightarrow\mbox{\boldmath$\nu_\beta$};t)
 \approx
\frac{\sin^{\2}{[2\theta]}}{2}
\left\{1 - \left[1 + \mbox{\sc Sp}^{\2}(t) \right]^{-\frac{1}{4}}
\exp{\left[-\frac{(\Delta \mbox{v} \, t)^{\2}}{2a^{\2}\left[1 + \mbox{\sc Sp}^{\2}(t)\right]}\right]}
\cos{\left[\Delta E \, t + \Theta(t)\right]}
  \right\}
\label{25A}
\end{equation}
and illustrated by Fig.~\ref{an8}
Obviously, the larger is the value of $a \, \bar{E}$, the smaller are the wave-packet effects.
If it was sufficiently large to not consider the second order corrections of Eq.~(\ref{11}),
we could rewrite the probability only with the leading terms ({\em slippage} effect),  
\begin{eqnarray}
&& P(\mbox{\boldmath$\nu_\alpha$}\rightarrow\mbox{\boldmath$\nu_\beta$};t) \approx
\frac{\sin^{\2}{[2\theta]}}{2}
 \left\{1- \exp{\left[-\frac{(\Delta \mbox{v} \, t)^{\2}}{2\, a^{\2}}\right]}\cos{[\Delta E \, t ]}\right\}
\label{20AA}
\end{eqnarray}
which corresponds to the same result obtained by \cite{DeL04}.
By assuming an UR propagation regime with $t \approx L$ and $E_i \sim p_{\0}$,
under {\em minimal slippage} conditions ($\Delta \mbox{v} \, L \ll a$), the Eq.(\ref{20AA}) reproduces the {\em standard} plane wave result,
\begin{eqnarray}
P(\mbox{\boldmath$\nu_\alpha$}\rightarrow\mbox{\boldmath$\nu_\beta$};L) 
&\approx& 
\frac{\sin^{\2}{[2\theta]}}{2}\left\{1- \cos{\left[\frac{\Delta m^{\2}}{2 p_{\0}} \, L \right]}\right\},
\label{20A}
\end{eqnarray}
since we have assumed $a \, \bar{E} \gg 1$.

By summarizing, we have obtained an explicit expression for the flavor conversion formula
for (U)R and NR propagation regimes which is valid under the particular assumption of a sharply peaked momentum
distribution. 
We have also observed that the {\em spreading} represents a minor
modification effect which is practically irrelevant for (ultra)relativistic propagating particles.
In particular, the {\em intermediate} wave-packet prescription elaborated here
can be discussed in the context of neutrino flavor oscillations.
We have concentrated our arguments on the existence of an additional time-dependent phase in the
oscillating term of the flavor conversion formula.
Such an additional phase presents an analytic dependence on time
which changes the oscillating character in a peculiar way. These modifications are minimal 
when $p_{\0}^{\2} \approx \frac{1}{3}\bar{E}^{\2}$ and more relevant for NR propagation regimes.
The existence of an additional time-dependent phase in the
oscillating term of the flavor conversion formula
coupled with the modified {\em spreading} effect can represent some minor but accurate
modifications to the (ultra)relativistic oscillation probability formula which leads to important
corrections to the phenomenological analysis for obtaining accurate ranges and limits for the 
neutrino oscillation parameters.
The relevance of such second-order corrections depends essentially on the value of the product between the
wave=packet width $a$ and the averaged energy flux $\bar{E}$ which parameterize the power series 
expansion here proposed and quantified.
 
Finally, we know the necessity of a more sophisticated approach
is understood. It involves a field-theoretical treatment.
Derivations of the oscillation formula resorting to field-theoretical methods are not very
popular. They are thought to be very complicated and the existing quantum field computations of the
oscillation formula do not agree in all respects \cite{Beu03}.
The  Blasone and Vitiello (BV) model \cite{Bla95,Bla03} to neutrino/particle mixing and oscillations
seems to be the most distinguished trying to this aim.
They have attempt to define a Fock space of weak eigenstates to derive a nonperturbative oscillation formula.
Also with Dirac wave-packets, the flavor conversion formula can be reproduced \cite{Ber04B}
with the same mathematical structure as those obtained in the BV model \cite{Bla95,Bla03}.
In fact, both frameworks deserve a more careful investigation since
the numerous conceptual difficulties hidden in the quantum oscillation phenomena still
represent an intriguing challenge for physicists.

\acknowledgments
This work was supported by FAPESP (PD-04/13770-0).

\end{document}